\documentclass[11pt]{cernrep}
\usepackage{graphicx}
\usepackage{here}

\def\NJ{N_{\J}}
\def\J{J/\psi}
\def\Nc{N_c}
\def\Ncbar{N_{\overline c}}
\def\ccbar{c \overline c}
\def\Nccbar{N_{\ccbar}}
\begin{document}
\title{QUARKONIUM FORMATION FROM UNCORRELATED QUARK-ANTIQUARK PAIRS} 
\author{R. L. Thews}
\institute{Department of Physics, University of Arizona, Tucson, AZ 
USA}
\maketitle
\begin{abstract}
The goal of this section is to assess the possibility that
quarkonium  production rates
may be enhanced in nucleus-nucleus interactions at the LHC
relative to that predicted by extrapolation of processes thought
to be dominant at lower energy.  This enhancement could follow from the
effects of incoherent
recombination mechanisms involving uncorrelated pairs of heavy
quarks and antiquarks which result from multiple pair production.
Two different approaches have been considered:
statistical hadronization and kinetic formation. Updated predictions
relevant to Pb+Pb collisions at the LHC are given.
\end{abstract}
\section{INTRODUCTION}
\label{thews.intro}
The utility of heavy quarkonium
production rates in nuclear collisions as
a signature of color deconfinement
was proposed more than 15 years ago \cite{Matsui:1986dk}.
Since one expects that the long-range color confining potential will be
screened in a deconfined medium, the quark and antiquark constituents of
bound states will be liberated.  As the system expands and cools, these
constituents will, in general, diffuse away from each other
to separations  larger than typical hadronic dimensions.
When the confining potential reappears, a given heavy
quark will not be able to ``find'' its heavy antiquark partner and form heavy
quarkonium.  It must then bind with one of the antiquarks within range
at hadronization.  Since these antiquarks are predominantly the lighter
$u$, $d$, and $s$ flavors, the final hadronic states will preferentially be
those with ``open'' heavy flavor.  The result will be a
decreased population of heavy quarkonium relative to that which would
have formed if a region of deconfinement had not been present.  This
scenario as applied to the charm sector is known as $\J$ suppression.

At LHC energy, perturbative QCD estimates predict that
hundreds of pairs of charm-anticharm quarks will be produced
in a central lead-lead collision.
This situation provides a ``loophole'' in the Matsui-Satz
argument since there will be  copious numbers of heavy antiquarks in the
interaction region with which any given heavy quark may combine.  
In order for this to
happen, however, one must invoke a physical situation in which
quarkonium states can be formed from {\em all combinations}
of heavy quarks and antiquarks.
This of course
would be expected to be valid in the case that a space-time region
of color deconfinement is present but it is not necessarily limited to
this possibility.  

One can make a model-independent estimate of how
such a ``recombination" mechanism would depend on nuclear collision
observables. For a given charm quark,
the probability $\cal{P}$ to form a $\J$ is
proportional
to the number of available anticharm quarks relative to the number of
light antiquarks,
\begin{equation}
{\cal{P}} \propto \frac{\Ncbar}{N_{\overline u, \overline d, \overline s}}
\propto \frac{\Nccbar}{N_{\rm ch}} \, \, .
\end{equation}
In the second step, we have replaced the number of available anticharm
quarks by the total number of pairs initially produced, assuming that
the total number of bound states formed remains a small fraction of the total
$c \overline c$ production.
We normalize the number of light antiquarks by the number of
produced charged hadrons.
Since this probability  is generally very small, one can simply multiply by the
total number of charm quarks, $\Nc$, to obtain the number of
$\J$ expected in a given event,
\begin{equation}
\NJ \propto \frac{{\Nccbar}^2}{N_{\rm ch}} \, \, ,
\label{eqquadratic}
\end{equation}
where the use of the initial values $\Nccbar = \Nc = \Ncbar$ is again justified
by the relatively small number of bound states formed.

The essential property of this result is that the growth of $\NJ$, quadratic
in total charm, with energy \cite{Gavai:1994gb}
is expected to be much faster than
the growth of total particle production in heavy ion collisions
\cite{Bazilevsky:2002fz}.   Without this quadratic mechanism, $\J$ production 
is typically some small energy-independent fraction of total initial charm
production \cite{Gavai:1994in}.  We thus anticipate that the quadratic
formation will become dominant at sufficiently high energy.
Generic estimates of the significance of this type of formation
process can be made \cite{Thews:2001hy}.  Here we
look at specific predictions of two models which share the above
properties and update the expectations to LHC energies.

\section{Statistical Hadronization}
\label{thews.stat}

The statistical hadronization model is 
motivated by the successful fits of relative abundances
of light hadrons produced in high energy heavy ion interactions
according to a hadron gas in chemical and thermal equilibrium
\cite{Braun-Munzinger:2001ip}.  Extension of the model 
to hadrons containing heavy quarks
underpredicts the observed abundances.  This effect may be
attributed to the long time scales associated with
thermal production and annihilation of heavy quarks.
The statistical hadronization model as first formulated 
for charm quarks \cite{Braun-Munzinger:2000px} 
assumes that the $c \overline c$ pairs produced in the initial hadronic
interactions survive until their subsequent hadronization, at which time they
are distributed into hadrons according
to the same thermal equilibrium parameters that fit the light
hadron abundances.  Chemical equilibrium abundances are 
adjusted by a factor $\gamma_c$ which accounts for 
the non-thermal 
heavy quark density. One power of this factor multiplies a given 
thermal hadron population for each heavy quark or antiquark
contained in the hadron.  Thus the relative abundance of the
$\J$  to that of $D$ mesons, for example, may be enhanced in this model.

The value of $\gamma_c$ is determined by conservation of the heavy quark
flavor.
For the charm sector, the conservation constraint relates
the number of initially-produced c-cbar pairs $\Nccbar$ to their
distribution into open and hidden charm hadrons,
\begin{equation}
\Nccbar = {1\over 2}\gamma_c N_{\rm open} + {\gamma_c}^2 N_{\rm hidden},
\label{eqgcstat}
\end{equation}
where $N_{\rm open}$ is the number of hadrons containing one $c$ or
$\overline c$ quark and $N_{\rm hidden}$ is the number of hadrons containing
a $c \overline c$ pair. 
For most applications, $N_{\rm hidden}$
(and also multi-charm hadrons) can be neglected compared with
$N_{\rm open}$ due to the mass differences.  Thus the charm
enhancement factor is simply
\begin{equation}
\gamma_c = {2\Nccbar\over N_{\rm open}},
\end{equation}
leading directly to the quadratic dependence of the hidden charm hadron
population on $\Nccbar$.
One can then express the total number of $\J$ in terms of
the various thermal densities, $n_i$, and the total number of $c \overline
c$ pairs, $\Nccbar$.
One factor of system volume $V$ remains implicit here.  It is generally
replaced by the ratio of number to density for total charged
hadrons, $n_{\rm ch}/N_{\rm ch}$. Then the number of $\J$ produced obeys the
generic form anticipated in Eq.~(\ref{eqquadratic}).
\begin{equation}
\NJ = 4 {n_{\rm ch} n_{\J}\over {n_{\rm open}}^2} {\Nccbar^2\over N_{\rm ch}}
\label{eqstathad}
\end{equation}                                                                 

For collider experiments such as those at the LHC and RHIC, 
relating the corresponding central rapidity densities will be
more relevant.  Since Eq.~(\ref{eqstathad}) is homogeneous in
the total particle and
quark pair numbers, it will also be valid if these 
are replaced by their rapidity densities.  To get an order of
magnitude estimate,    
we choose a ``standard'' set of thermal
parameters, $T = 170$ MeV and $\mu_B \approx 0$, for which the thermal
density ratio is approximately $0.5$.  For a specific normalization,
we assume $dN_{\rm ch}/dy$ = 2000 for a central
collision at the LHC and take the initial charm rapidity density to be 
$dN_{c \overline{c}}/dy = 25$, roughly corresponding to $\Nccbar = 200$ 
for central collisions $(b=0)$.  
Using these inputs, one predicts 
$d\NJ/dy = 0.625$, indicating that several $\J$ will form through 
statistical hadronization in a  central collision.  
To put this number in perspective,
it is revealing to form the $\J$ to $\Nccbar$ rapidity
density ratio, 0.025 with the same assumptions.
For comparison, one expects the corresponding hadronic production ratio 
to be of $0.01$.  This number would then be significantly reduced
if placed in a region of color deconfinement.  Thus the efficiency of
$\J$ formation via statistical hadronization at the LHC is expected
to be substantial. 

These numbers can be easily adjusted to other charm and
charged particle densities using Eq.~(\ref{eqstathad}).  
Variations of the thermal parameters can also be investigated.
For example, if the hadronization
temperature is decreased to 150 MeV, the prefactor
combination of thermal densities increases by approximately
a factor of two.  

The centrality dependence is controlled by the behavior
of $\Nccbar$ and $N_{\rm ch}$.  The former should be 
proportional to the nuclear overlap function $T_{AA}(b)$ but is 
generally recast in terms of the dependence on the
number of nucleon participants, $N_{\rm part}$.  The calculation of 
$N_{\rm part}$ requires a model
calculation dependent on the total inelastic cross section, $\sigma_{\rm in}$, 
as well as the nuclear geometry.  We parameterize the 
expected behavior as a power-law $\Nccbar \propto 
{N_{\rm part}}^{4/3}$.  However, there will be deviations from this
behavior for the larger values of $\sigma_{\rm in}$
expected at the LHC \cite{Andronic:2002pj}.  The centrality dependence
of $N_{\rm ch}$ at RHIC is also consistent with a power-law with
exponent $\approx$ 1.2
\cite{Bazilevsky:2002fz,Kharzeev:2001yq}.  We will use the same dependence for
our estimates at the LHC.  
It is clear that for sufficiently peripheral collisions one will encounter
situations in which the average number of initially produced $c \overline c$
pairs is of order unity or less.  At this point, one must revisit 
the assumptions of the original statistical
hadronization model which assumed a grand canonical ensemble.  The grand
canonical approach is valid only when $\Nccbar$ is large enough for the
fluctuations about the average value to be neglibile.   Thus for
peripheral collisions, one must recalculate the statistical results
in the canonical approach where the charm number is exactly 
conserved, as noted in \cite{Gorenstein:2000ck}.  Charm conservation can be
implemented via a correction factor \cite{Cleymans:1990mn},
\begin{equation}
\Nccbar = {1\over 2} \gamma_c N_{\rm open} {{I_1(\gamma_c N_{\rm open})}\over 
{I_0(\gamma_c N_{\rm open})}} + {\gamma_c}^2 N_{\rm hidden}.
\end{equation} 
In the limit of large $\gamma_c N_{\rm open}$, 
the ratio of Bessel functions approaches unity and the grand canonical result
is recovered.  In the opposite limit when
$\gamma_c N_{\rm open} \rightarrow 0$, the ratio of Bessel functions approaches
${1\over 2} \gamma_c N_{\rm open}$.  In this limit, the dependence on
$\Nccbar$ in Eq.~(\ref{eqstathad}) changes from quadratic to linear.
At the LHC this effect will not be relevant until one reaches
very peripheral events, but at lower energies it can be 
significant over a much larger range of centralities \cite{Gorenstein:2001xp}.
\begin{figure}[htb]
\begin{center}
\includegraphics[width=12.5cm]{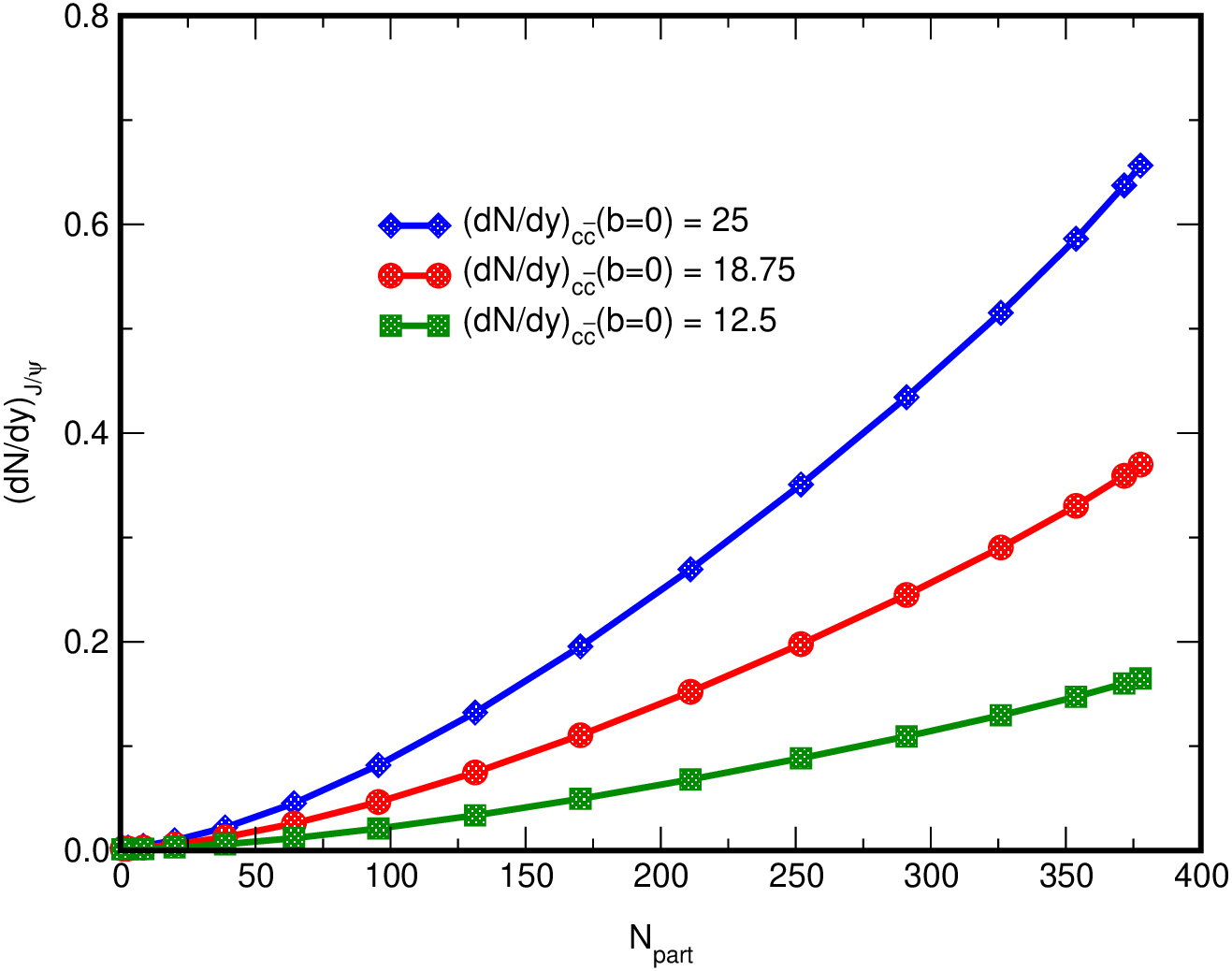}
\caption{Statistical hadronization results for $\J$ production as a function 
of $N_{\rm part}$ at the LHC.}
\label{lhcstatcharm}
\end{center}
\end{figure}

The results for $d\NJ/dy$ as a function of $N_{\rm part}$ at the LHC are 
shown in Fig.~\ref{lhcstatcharm}.
The results are shown for three different values of $dN_{\ccbar}/dy (b=0)$, 
corresponding to $\Nccbar(0) \approx 200$, 150, and 100.  
There is a rapid increase with
centrality due to the quadratic dependence of $\NJ$ on
$\Nccbar$. 

\begin{figure}[hbt]
\begin{center}
\includegraphics[width=12.5cm]{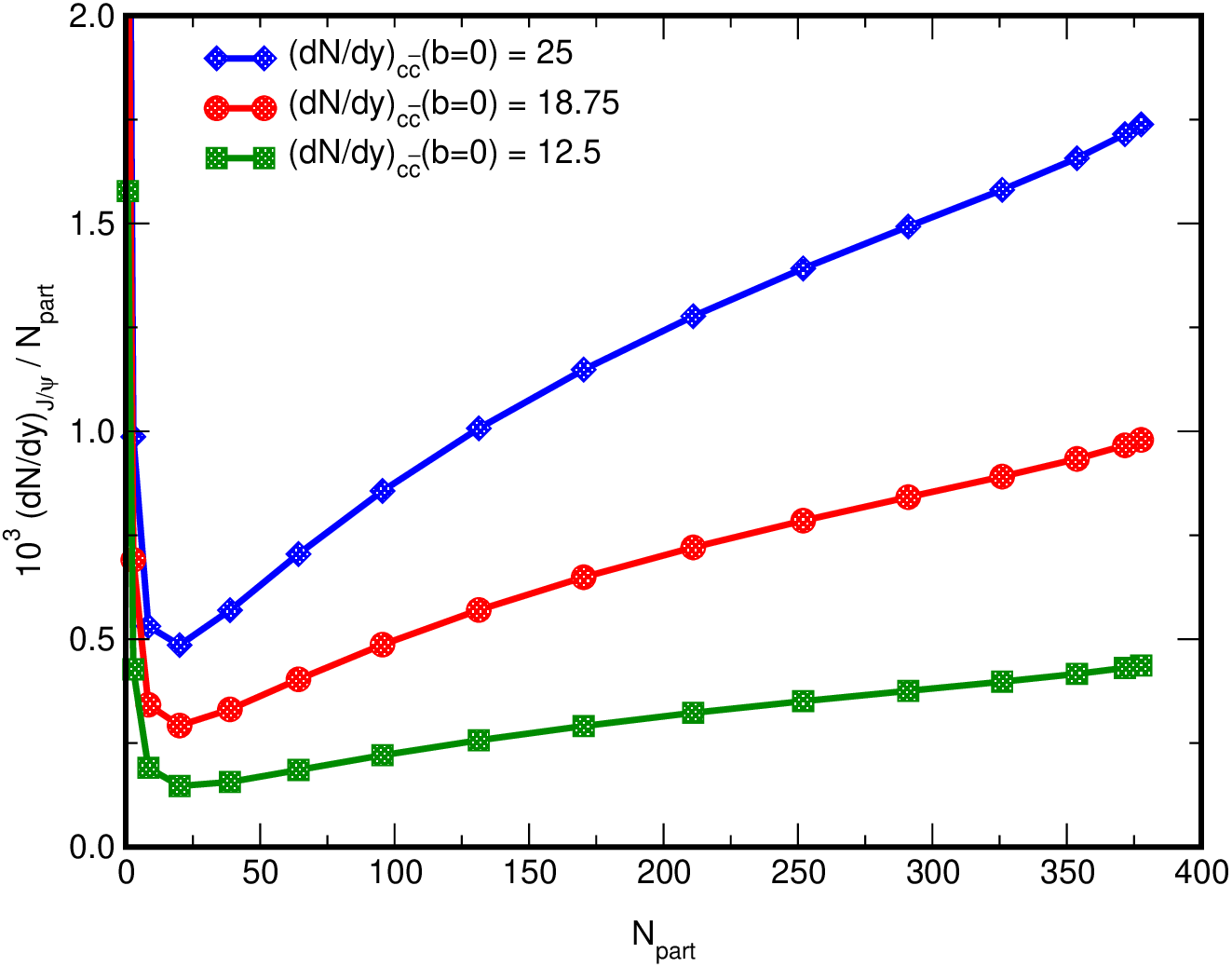}
\caption{The statistical hadronization results divided by $N_{\rm part}$
as a function of $N_{\rm part}$.}
\label{lhcstatcharmovernp}
\end{center}
\end{figure}

\begin{figure}[hbt]
\begin{center}
\includegraphics[width=12.5cm]{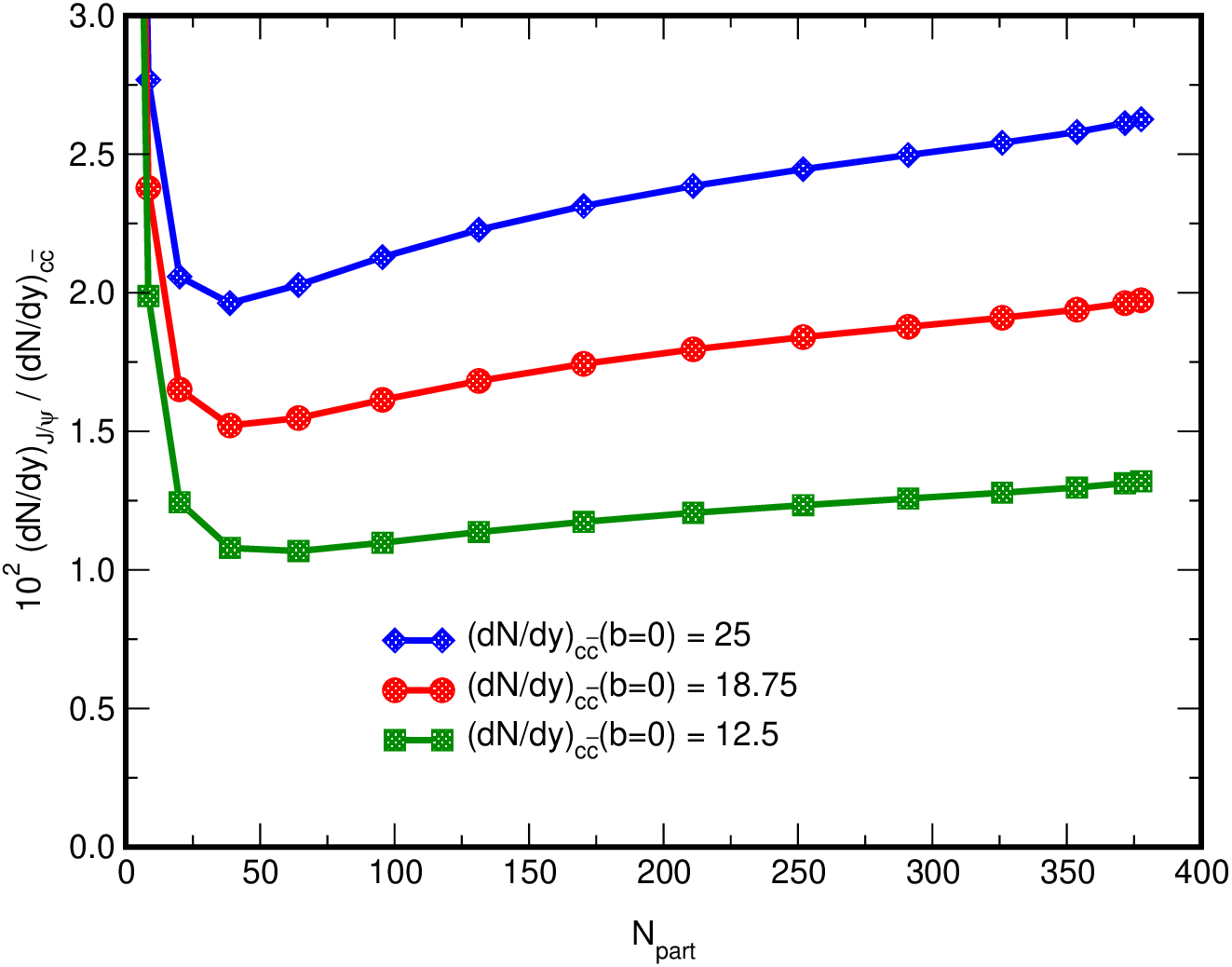}
\caption{The statistical hadronization results for $J/\psi$ production
at the LHC, divided by the open charm multiplicity, $d\Nccbar/dy$, as a
function of $N_{\rm part}$.}
\label{lhcstatcharmovercharm}
\end{center}
\end{figure}

It is also interesting to look at these results normalized by $N_{\rm part}$, 
shown in Fig.~\ref{lhcstatcharmovernp}.
This ratio also increases with centrality, providing a signature
for the statistical hadronization process that is less dependent on $\Nccbar$
for the overall normalization.   The corresponding
results when normalized by $d\Nccbar/dy$ are shown in 
Fig.~\ref{lhcstatcharmovercharm}.  The same general behavior is seen
but the increase with centrality is less pronounced since the
$d\Nccbar/dy$ is assumed to vary with a larger power, ${N_{\rm part}}^{4/3}$.
All of these ratios are at the percent level for central collisions and
hence are larger than expected if the total
$\J$ population were due to initial production followed by 
suppression in a deconfined medium.

The region of very peripheral collisions deserves some separate
comments.  First, there is a rise at low $N_{\rm part}$ in both 
Figs.~\ref{lhcstatcharmovernp} and \ref{lhcstatcharmovercharm} due 
to the onset of corrections from the canonical ensemble treatment.  However, 
the extremely large values of the ratios as $N_{\rm part} \rightarrow 0$ 
are an artifact of the decreasing
interaction volume, $V \rightarrow 0$.  
This calculation must be cut off before $N_{\rm part} = 2$, i.e. only one 
interacting pair. It is also in this region where one must 
take into account the remaining $\J$ from initial production.  Since
the survival probability is maximum for very peripheral collisions
and the statistical hadronization process is least effective in this
same region, there will be a crossover in the relative importance of
the two mechanisms.  Some studies have already been performed for this
situation at SPS and RHIC energies 
\cite{Grandchamp:2002iy,Grandchamp:2002wp,Grandchamp:2001pf}.

Finally, there is another lower cutoff in centrality for 
the statistical hadronization results, 
needed to avoid a contradiction with the $\psi\prime/(J/\psi)$ ratio at the
SPS.  Since both of these states
receive identical factors of $\gamma_c$, their ratio must be that
predicted for chemical equilibrium in the absence of any charm
enhancement or suppression.  Although the measured ratio appears
to be consistent for more central collisions
\cite{Sorge:1997bg}, there is an indication that it rises sharply for
more peripheral collisions.  Most treatments have
thus inserted a cutoff of $N_{\rm part}$ = 100, below which model predictions
become inconsistent \cite{Braun-Munzinger:2000px}.

The numerical values for $d\NJ/dy$ are tabulated as a function of
impact parameter in Table 1 for the
three choices of initial charm multiplicity density and the default values
of all other quantities.


\section{Kinetic Formation in a Deconfied Region}
\label{thews.kinetic}

The kinetic model has been developed \cite{Thews:2000rj,Thews:2001em}
to investigate the possibility that $\J$ may form directly in
a deconfined medium.  This formation takes advantage of the mobility
of the initially-produced charm quarks in a deconfined region.
In order to motivate this view, consider the ``standard'' physical picture
of deconfinement in which quarkonium is suppressed by collisions with free
gluons in the medium \cite{Kharzeev:1994pz}.  Then the formation process,
in which a $c$ and $\overline c$ in
a relative color octet state are captured into a color-singlet
quarkonium bound state and emit a color octet gluon,
is simply the
inverse of the the breakup reaction responsible for
the suppression.  This is an inevitable consequence of the suppression picture.

The proper time evolution of the $\J$ population
is given by the rate equation
\begin{equation}\label{eqkin}
\frac{d\NJ}{d\tau}=
  \lambda_{\mathrm{F}} \frac{N_c\, N_{\overline c}}{V(\tau)} -
    \lambda_{\mathrm{D}} \NJ\, \rho_g\, \, ,
\end{equation}
where $\rho_g$ is the gluon number density
and $V(\tau)$ is the time-dependent volume of the deconfined spatial region.
The reactivities $\lambda_{\rm F,D}$ are
the reaction rates, $\langle \sigma v_{\mathrm{rel}} \rangle$,
averaged over the momentum distributions of the initial
participants, i.e. $c$ and $\overline c$ for $\lambda_{\rm F}$ and
$\J$ and $g$ for $\lambda_{\rm D}$.                             

The solution of Eq.~(\ref{eqkin}) grows quadratically
with $\Nccbar$, as long as $\NJ \ll \Nccbar$.  In
this case, we have
\begin{equation}
\NJ(\tau_f) = \epsilon(\tau_f) \left[\NJ(\tau_0) +
\Nccbar^2 \int_{\tau_0}^{\tau_f} \, d\tau
\lambda_{\mathrm{F}}\, [V(\tau)\, \epsilon(\tau)]^{-1} \right] \, \, .
\label{eqbeta}
\end{equation}
The function $\epsilon(\tau_f) =
\exp(-\int_{\tau_0}^{\tau_f} d\tau \lambda_{\mathrm{D}}\, \rho_g)$
would be the suppression factor if formation were neglected.  
                                                                        
The quadratic factor $\Nccbar^2$ is present, as expected, for
the additional formation process.  The
normalization factor of $N_{\rm ch}$ is not immediately evident, but
is implicit in the system volume factor.
This volume is now time-dependent, accounting for the decreasing charm quark
density during expansion.  
Here the transverse area of the deconfined region is determined not just by the
nuclear geometry but by the dynamics which determine the extent of the 
deconfined region.  This area is modeled by the energy density in terms of 
the local participant density in the transverse plane, $n_{\rm part}(b, s=0)$. 
The transverse area is defined by the ratio of the participant number 
to the local participant density.  Note that the maximum local density is
at $s=0$.  Thus,
\begin{equation}
A_T(b) = A_T(0) [N_{\rm part}(b) 
n_{\rm part}(0, s=0)/N_{\rm part}(0) n_{\rm part}(b, s=0)]
\end{equation}                                      
These area effects will be more explicit when the centrality
dependence is considered.    

The numerical results depend on a number of parameters, including the
initial volume and temperature, the time expansion profile, the
reaction cross sections, the behavior of the quarkonium 
masses and binding energies in the deconfined region, and the charm quark
momentum distributions.  For specifics, see Ref. \cite{Thews:2002jg}.
Our previous results have used initial values $\Nccbar = 200$, 150, and 
100, spanning a reasonable range of expectations \cite{Vogt:2001nh}.
The results are very sensitive to the initial charm quark momentum 
distributions, as may be expected. We assume the charm $p_T$ distributions
are gaussian and the charm rapidity distributions are flat 
over a plateau of variable 
width, $\Delta y$.  The range $1 < \Delta y < 7$ spans the range between an 
approximate thermal momentum distribution, $\Delta y \approx 1$, to a 
distribution similar to that of the initial pQCD production, $\Delta
y \approx 7$.

\begin{figure}[htb]
\begin{center}
\includegraphics[width=12.5cm,clip]{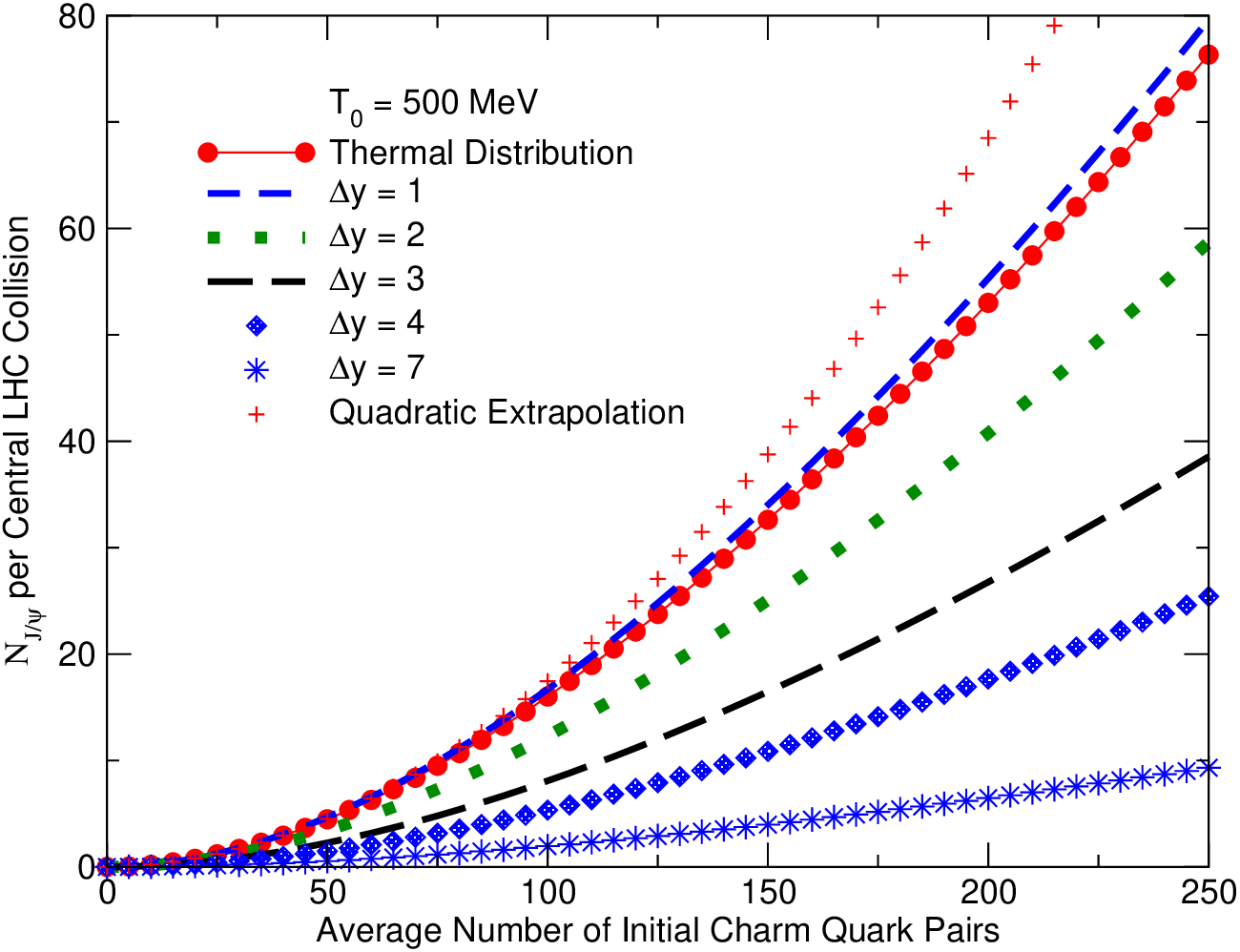}
\caption{The $J/\psi$ production per central LHC collision in the kinetic
model as a function of the initial number of $c \overline c$ pairs.}
\label{lhcbzero}
\end{center}
\end{figure}

The results as a function of the initial number of $c \overline c$ pairs 
produced in central collisions are shown in Fig.~\ref{lhcbzero}.
There is a rapid decrease in formation with increasing $\Delta y$.
The quadratic dependence on $\Nccbar$ is evident, but there is also 
a substantial linear component in some of the curves.  This linear contribution
arises because the final $\J$ formation by this mechanism
is large enough for exact charm conservation to reduce the number of $c$
and $\overline c$ quarks available to participate
in the formation process.  The curve labeled
``Quadratic Extrapolation'' uses a quadratic
dependence derived from a fit 
valid only for low $\Nccbar$.  Note that the result for a
thermal distribution is very similar to the assumption $\Delta y =1$.

\begin{figure}[htb]
\begin{center}
\includegraphics[width=12.5cm,clip]{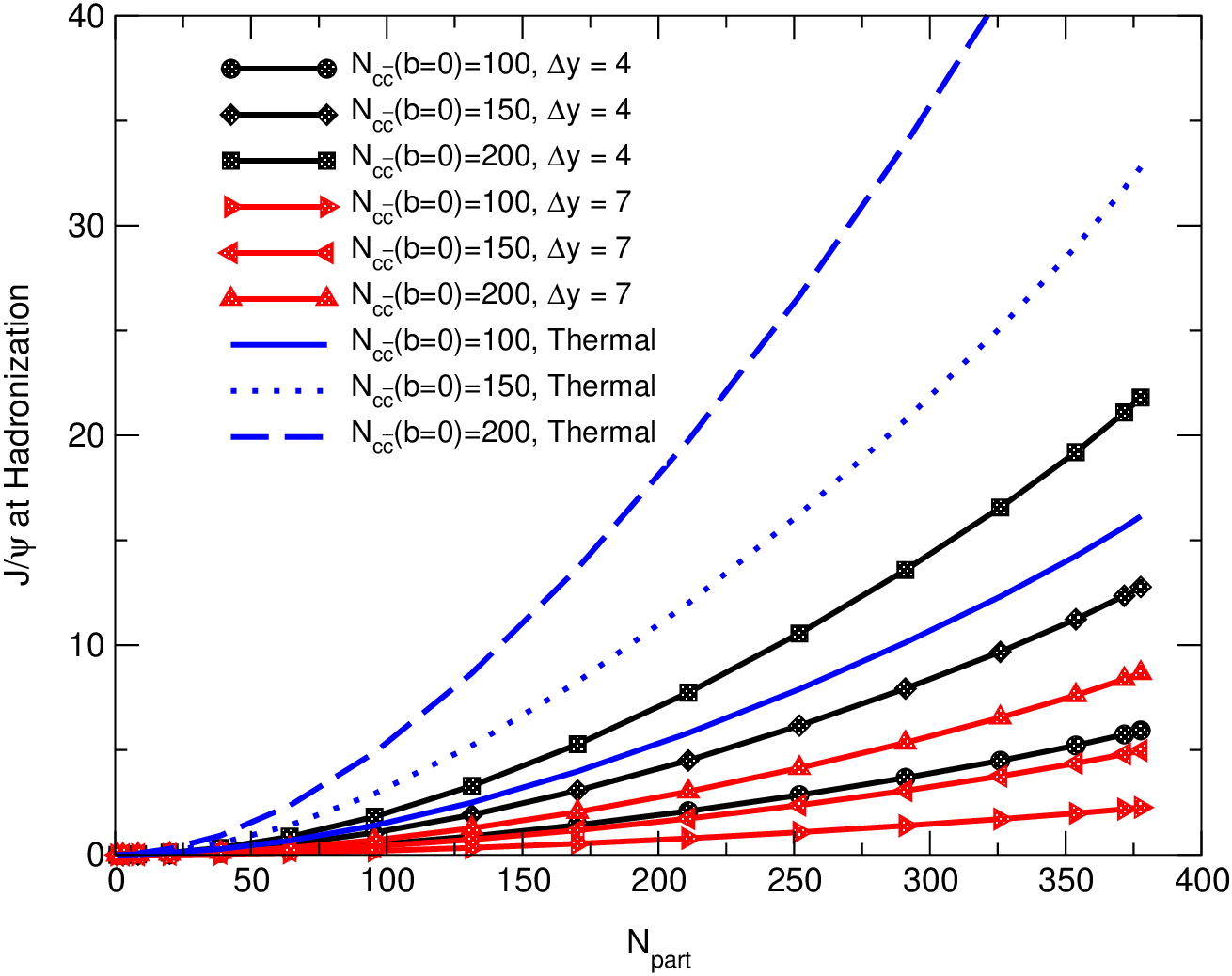}
\caption{The centrality dependence of $J/\psi$ production in the kinetic
model.}
\label{lhcb}
\end{center}
\end{figure}

The corresponding centrality dependence is presented in 
Fig.~\ref{lhcb}, where we give $\NJ$ at hadronization
for three different initial charm quark momentum distributions, thermal,
$\Delta y = 4$ and $\Delta y = 7$, as well as for our three choices of 
$\Nccbar(b=0)$.  

\begin{figure}[h]
\begin{center}
\includegraphics[width=12.5cm,clip]{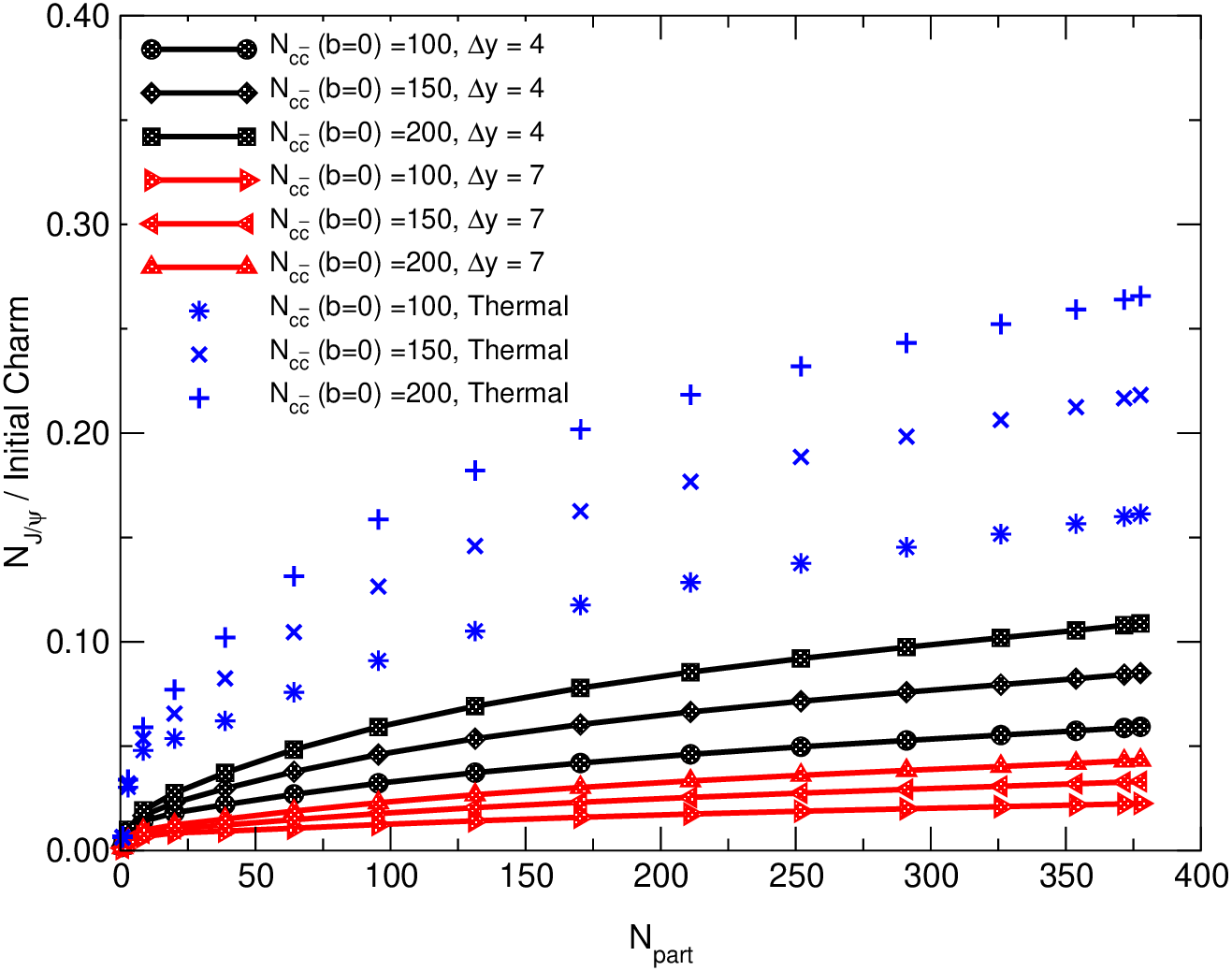}
\caption{The ratio of the number of produced $J/\psi$'s in the kinetic model
to the initial number of $c \overline c$ pairs as a function of $N_{\rm
part}$.} 
\label{lhcjpsiovercharm}
\end{center}
\end{figure}

Finally, the ratio of final $\J$ to initial charm production is shown
in Fig. \ref{lhcjpsiovercharm} using 
the same parameters in Fig.~\ref{lhcbzero}.  These ratios are most
easily compared to either initial production or
suppression.  There is a substantial variation
in the predictions and it is evident that a simultaneous
measurement of open charm will be required for an
interpretation.  However, the centrality dependence is
opposite to that expected in any pure suppression scenario.

\begin{table}[htb]
\begin{center}
\caption{Comparison of $J/\psi$ production at the LHC by the statistical
hadronization (left-hand side) and kinetic formation (right-hand side) models.}
\label{thews.tab1}
\begin{tabular}{|c|c|c|c||c|c|c|} \hline \hline
 & \multicolumn{3}{|c||}{$dN_{\J}/dy$ (Statistical)} &
\multicolumn{3}{|c|}{$N_{\J}$ (Kinetic, LO Charm)}\\ 
 & \multicolumn{3}{|c||}{$dN_{c\overline c}(0)/dy$} &
\multicolumn{3}{|c|}{$N_{c \overline c}(0)$} \\ \hline \hline
$b$ (fm) & 25 & 18.75 & 12.5 & 200 & 150 & 100 \\ \hline \hline
0&0.656&0.370&0.165&4.0&2.26&1.03 \\ \hline
1&0.637&0.359&0.160&3.85&2.19&1.00\\ \hline
2&0.586&0.330&0.147&3.51&2.00&0.91\\ \hline
3&0.515&0.290&0.130&3.04&1.73&0.79\\ \hline
4&0.434&0.245&0.109&2.50&1.43&0.65\\ \hline
5&0.351&0.198&0.088&1.97&1.12&0.51\\ \hline
6&0.270&0.152&0.068&1.46&0.84&0.38\\ \hline
7&0.196&0.110&0.050&1.01&0.58&0.27\\ \hline
8&0.132&0.075&0.034&0.65&0.38&0.18\\ \hline
9&0.082&0.046&0.021&0.38&0.22&0.10\\ \hline
10&0.045&0.026&0.012&0.20&0.12&0.057 \\ \hline
11&0.022&0.013&0.0061&0.087&0.054&0.028 \\ \hline
12&0.0097&0.0058&0.0029&0.034&0.022&0.012 \\ \hline
13&0.0045&0.0029&0.0016&0.011&0.0075&0.0041 \\ \hline
14&0.0028&0.0019&0.0012&0.0021&0.0013&$6.8 \times 10^{-4}$ \\ \hline
15&0.0025&0.0018&0.0012&$1.8 \times 10^{-4}$&$1.0 \times 10^{-4}$ & $5.1 \times
10^{-5}$\\ \hline \hline
\end{tabular}
\end{center}

\end{table}

We have updated the calculations to include the
charm quark momentum distribution from a leading order pQCD calculation
\cite{thewsinprep}.  The rapidity distribution has a somewhat larger 
effective $\Delta y$ and the $p_T$ distribution does not fall as fast 
as a simple gaussian.  As a result, the formation efficiency is further 
reduced.  Such distributions may be most relevant, given
preliminary results from RHIC \cite{Adcox:2002uc,Frawley}.

\begin{table}[htb]
\begin{center}
\caption{Kinetic $J/\psi$ formation at the LHC assuming both
thermal charm momentum (left-hand side) and
$\Delta y = 4$ (right-hand side).}
\label{thews.tab2}
\begin{tabular}{|c|c|c|c||c|c|c|} \hline \hline
 &\multicolumn{3}{|c||}{$N_{\J}$ (Thermal)}&
\multicolumn{3}{|c|}{$N_{\J}$ ($\Delta y = 4$)} \\ 
 & \multicolumn{3}{|c||}{$N_{c \overline c}(0)$} &
\multicolumn{3}{|c|}{$N_{c \overline c}(0)$} \\ \hline \hline
$b$ (fm) & 200 & 150 & 100 & 200 & 150 & 100 \\ \hline \hline
0&52.7&32.5&16.4&17.5&10.8&5.48\\ \hline
1&50.5&31.2&15.8&16.8&10.4&5.25\\ \hline
2&44.8&27.7&14.0&14.9&9.21&4.65\\ \hline
3&37.0&22.9&11.5&12.3&7.62&3.82\\ \hline
4&28.6&17.7&8.73&9.54&5.89&2.90\\ \hline
5&20.7&12.7&6.05&6.90&4.23&2.01\\ \hline
6&13.8&8.32&3.72&4.61&2.77&1.24\\ \hline
7&8.28&4.71&2.14&2.76&1.57&0.71\\ \hline
8&4.10&2.36&1.10&1.36&0.79&0.37\\ \hline
9&1.78&1.04&0.50&0.59&0.35&0.16\\ \hline
10&0.65&0.39&0.19&0.22&0.13&0.064\\ \hline
11&0.19&0.12&0.063&0.065&0.040&0.021\\ \hline
12&0.048&0.032&0.018&0.016&0.010&0.006\\ \hline
13&0.011&0.0078&0.0049&0.0037&0.0026&0.0016\\ \hline
14&0.0026&0.0019&0.0012&$8.6 \times 10^{-4}$&$6.3 \times 10^{-4}$&
$4.1 \times 10^{-4}$ \\ \hline
15&$5.9 \times 10^{-4}$ &$4.4 \times 10^{-4}$ & $2.9 \times 10^{-4}$&
$2.0 \times 10^{-4}$& $1.5 \times 10^{-4}$&$9.7 \times 10^{-5}$\\ \hline \hline
\end{tabular}
\end{center}
\end{table}

The numerical values for $\NJ$ are compared with the statistical hadronization 
model results for $dN_{\J}/dy$ in Table~\ref{thews.tab1}.
The overall magnitudes are comparable, although
the centrality dependences differ somewhat.  Thus details such as the 
resulting $\J$ momentum distributions will be required to differentiate 
between these two models \cite{thewsinprep}.
For completeness, $\NJ$ for the thermal distributions and the assumption 
$\Delta y = 4$ are presented in Table~\ref{thews.tab2}.

\section{Conclusions}
\label{thews.conclus}

The "smoking gun" signature of the quarkonium formation mechanism
is the quadratic dependence on total charm.  For central
collisions at the LHC one
expects that this feature will lead to a total $\J$ rate greater than
that produced by an incoherent superposition
of the initial nucleon-nucleon collisions, even without any subsequent
suppression due to deconfinement effects.  
In addition, the centrality dependence can be used to identify the 
quadratic dependence on charm assuming that the initial charm
production scales with the number of binary collisions.  Binary scaling 
leads to an 
increase of the ratio of $\J$ to initial charm as the collision
centrality increases, independent of specific parameters which control
the overall magnitudes.  A simultaneous measurement
of total charm will be essential for such conclusions to be drawn.

Uncertainties in the absolute magnitude of the formation process are
inherent in the model parameters.  For statistical hadronization,
one can constrain the thermal parameters to within a factor of
two using the observed hadron populations.  There is some additional
uncertainty related to the lower cutoff on centrality needed to ensure 
the quarkonium ratios are consistent with an overall thermal picture.  There 
is also the possibility that the correction for canonical ensemble effects 
will involve a thermal volume parameter not necessarily equal to the
total system volume \cite{Redlich:2001kb}.  In addition, the formation 
mechanism could be limited to those charm quarks whose phase space separation 
is within some maximum value, introducing another as yet
unconstrained parameter \cite{Grandchamp:2002wp}.  With kinetic 
formation, a similar set of uncertainties exist. 
There are uncertainties in the space-time properties of the deconfinement
region. In addition, possible variations of charmonium binding energies 
and reaction cross sections in a deconfined region are at present not well
understood.  There are indications that the efficiency of the 
formation mechanism is considerably reduced when included in a partonic
transport calculation \cite{Zhang:2002ug}.  

The primary uncertainty in both models is still the initial number of charm
quarks and their momentum distributions.  The tabulated
$\J$ results should be regarded in this
light.  Thus numbers may be only an order of magnitude estimate.  
However, the variation with centrality 
and total initial charm 
should provide experimental signatures which are largely independent
of the overall magnitudes.

\section{Acknowledgments}

My thanks to Anton Andronic for discussions on the Statistical Model and
Martin Schroedter for updates on the Kinetic Model calculations.
This work was supported in part by U.S. Department of Energy 
Grant DE-FG03-95ER40937.

\end{document}